\documentclass[12pt,preprint]{aastex}

\begin{document}
\title{The Onfp Class in the Magellanic Clouds}

\author{Nolan~R.\ Walborn,\altaffilmark{1,8} 
Ian~D.\ Howarth,\altaffilmark{2,4} Christopher~J. Evans,\altaffilmark{3,4,6}  
Paul~A.\ Crowther,\altaffilmark{5,6} Anthony F.~J.\ Moffat,\altaffilmark{7,8} 
Nicole St-Louis,\altaffilmark{7} Cecilia Fari\~na,\altaffilmark{9} 
Guillermo L. Bosch,\altaffilmark{9,12}\\ Nidia~I.\ Morrell,\altaffilmark{10} Rodolfo~H.\ Barb\'a,$^{11,12}$ and Jacco Th.\ van Loon\altaffilmark{13}\\
~\\
~\\
~\\
~\\}

\altaffiltext{1}{Space Telescope Science Institute,* 3700 San Martin Drive, Baltimore, MD 21218; walborn@stsci.edu}
\altaffiltext{2}{Department of Physics and Astronomy, University College London, Gower Street, London WC1E 6BT, UK; idh@star.ucl.ac.uk}
\altaffiltext{3}{UK Astronomy Technology Centre, Royal Observatory Edinburgh, Blackford Hill, Edinburgh EH9 3HJ, UK; chris.evans@stfc.ac.uk}
\altaffiltext{4}{Visiting Astronomer, Anglo-Australian Observatory}
\altaffiltext{5}{Department of Physics and Astronomy, University of Sheffield, 
Hounsfield Road, Sheffield S3 7RH, UK; Paul.Crowther@sheffield.ac.uk}
\altaffiltext{6}{Visiting Astronomer, European Southern Observatory}
\altaffiltext{7}{D\'epartement de Physique, Universit\'e de Montreal, C.P.\ 6128, Succ.\ Centre-Ville, Montreal, QC H3C 3J7, Canada; 
moffat@astro.umontreal.ca, stlouis@astro.umontreal.ca}
\altaffiltext{8}{Visiting Astronomer, Cerro Tololo 
Inter-American Observatory, National Optical Astronomy Observatory, 
operated by the Association of Universities for Research in Astronomy, 
Inc., under a cooperative agreement with the NSF}
\altaffiltext{9}{Facultad de Ciencias Astron\'omicas y Geof\'{\i}sicas, Universidad Nacional de La Plata, 1900 La Plata, Argen\-tina and IALP-CONICET, Argentina; ceciliaf@fcaglp.unlp.edu.ar, guille@fcaglp.unlp.edu.ar}
\altaffiltext{10}{Las Campanas Observatory, Observatories of the Carnegie Institution of Washington, Casilla 601, La Serena, Chile; nmorrell@lco.cl}
\altaffiltext{11}{Departamento de F\'{\i}sica, Universidad de La Serena, Cisternas 1200 Norte, La Serena, Chile; and Instituto de Ciencias 
Astron\'omicas de la Tierra y del Espacio (ICATE--CONICET), Avenida 
Espa\~na 1512 Sur, J5402DSP, San Juan, Argentina; rbarba@dfuls.cl}
\altaffiltext{12}{Visiting Astronomer, Las Campanas Observatory, Observatories of the Carnegie Institution of Washington}
\altaffiltext{13}{Astrophysics Group, School of Physical \& Geographical Sciences, Keele University, Staffordshire ST5 5BG, UK; jacco@astro.keele.ac.uk}
\altaffiltext{*}{Operated by the Association of Universities for Research in Astronomy, Inc., under NASA contract NAS5-26555}

\begin{abstract}
The Onfp class of rotationally broadened, hot spectra was defined some
time ago in the Galaxy, where its membership to date numbers only eight.
The principal defining characteristic is a broad, centrally reversed
He~II $\lambda$4686 emission profile; other emission and absorption lines
are also rotationally broadened.  Recent surveys in the Magellanic Clouds 
(MCs) have brought the class membership there, including some related spectra, to 28.  We present a survey of the spectral morphology and rotational velocities, as a first step toward elucidating the nature of this class.  
Evolved, rapidly rotating hot stars are not expected theoretically, because 
the stellar winds should brake the rotation.  Luminosity classification of 
these spectra is not possible, because the principal criterion (He~II 
$\lambda$4686) is peculiar; however, the MCs provide reliable absolute 
magnitudes, which show that they span the entire range from dwarfs to 
supergiants.  The Onfp line-broadening distribution is distinct and shifted 
toward larger values from those of normal O dwarfs and supergiants with 
$>$99.99\% confidence.  All cases with multiple observations show 
line-profile variations, which even remove some objects from the class 
temporarily.  Some of them are spectroscopic binaries; it is possible that 
the peculiar profiles may have multiple causes among different objects.  
The origin and future of these stars are intriguing; for instance, they 
could be stellar mergers and/or gamma-ray-burst progenitors.    
\end{abstract}


\section{Introduction}

The Onfp designation was introduced by Walborn (1973) to describe several
peculiar (``p'') early-type spectra; the ``n'' is the classical descriptor 
of broadened absorption lines, generally ascribed to rapid rotation, 
while ``f'' denotes O-type spectra with He~II $\lambda$4686 and N~III 
$\lambda\lambda$4634--4640--4642 in selective (Walborn 2001) emission.
The defining peculiarity consists of a composite emission + absorption
profile at $\lambda$4686, which ranges among different objects from a 
strong absorption line with weak emission wings to a strong, broad emission 
line with a central absorption reversal.  The class was independently and 
simultaneously recognized by Conti \& Leep (1974), who designated 
it Oef, by implied analogy with the Be stars.

It is noteworthy that the two brightest Of stars in the sky, $\lambda$~Cephei 
(Walborn 1973, Conti \& Frost 1974) and $\zeta$~Puppis (Conti \& Niemela 1976), belong to this peculiar class.  Contrary to what might be expected from that datum, modern classifications of about 400 Galactic O stars have yielded only six additional members (Ma\'{\i}z Apell\'aniz et al.\ 2004).  Thus, they are rare objects.  The known Galactic Onfp spectra are listed in Table~1 with references.  

Extensive recent studies and surveys of the rich O-type population of the
Magellanic Clouds (MCs) have yielded an increasing sample of Onfp and related spectra, which provide the motivation and subject of this work.  A larger sample with known distances may provide vital information toward a physical interpretation of the class.  The number of objects available for
discussion here is 28.  However, it should not be inferred that they have a higher incidence in the MCs; e.g., Evans et al.\ (2004) found just one among 139 O stars in their study, and the current survey by I.D.H.\ comprises 191 O stars among which there are four new Onfp spectra. These objects are included in the present sample.

Mihalas \& Conti (1980) proposed that the phenomenology of $\zeta$~Pup might result from corotation of the inner stellar wind enforced by a weak (50~gauss) magnetic field.  From spectrophotometry of the variable H$\alpha$ profile, Moffat \& Michaud (1981) suggested that $\zeta$~Pup might be a magnetic oblique rotator with a period of about 5~d.  However, Donati, Howarth et al.\ (unpublished) were unable to detect a field on this star to a comparable limit, albeit orientation dependent.  Henrichs (1991) reported correlated $\lambda$4686 and UV wind-profile variations, possibly periodic, in
$\lambda$~Cephei on very short timescales, and he inferred that non-radial
pulsations may be present.  The similarity of the Onfp He~II $\lambda$4686 
profiles to the Balmer profiles of Be stars might suggest formation in a 
disk.  However, Bouret, Hillier, \& Lanz (2008) have successfully modeled 
the $\lambda$4686 and H$\alpha$ profiles of $\zeta$~Pup with a clumped, 
rotating wind alone.  We shall present available and suggest future 
observations that may eventually elucidate the origin(s) of the peculiar 
Onfp spectral morphology.

\section{Observations}

Information about the stars discussed here is presented in Table~2; the 
first three are in the Small Magellanic Cloud (SMC), and the remaining 25 in the Large Magellanic Cloud (LMC).  (There are two independent observations for two stars: Sk~190 = 2dFS~3975 in the SMC, and 2dFL~52--171 = AA$\Omega$~30Dor~142 in the LMC.)  In this section we provide observational parameters of the diverse data sources, with more detail for cases in which that information is not available in references listed in the table.  

Previously unpublished data of Crowther are from the UV-Visual Echelle 
Spectrograph (UVES) at the European Southern Observatory (ESO) Very Large Telescope (VLT).  They were obtained under program 074.D-0109 during 2004 November~23--30.  The slit width was 1.2\arcsec\ and the UVES blue arm with a central setting of 4370~\AA\ provided continuous coverage of 3730--5000~\AA\ at a resolving power of 37,500 on a single $2 \times 4$~K EEV CCD.  The observation of AV~80 in the SMC used here was also obtained during that run, although an earlier observation is referenced in Table~2.  The observation of Sk~$-67^{\circ}$~111 is from a prior UVES run as specified by Walborn et al.\ (2002a).

The VLT FLAMES/Giraffe data of Evans et al.\ (2006) have a resolving power
of 23,300 at He~II $\lambda$4686.

The 2-degree Field LMC (2dFL) survey by Howarth was conducted with the 2dF fiber-fed multiobject spectrographs on the 3.9~m Anglo-Australian Telescope; observations discussed here were obtained during the nights of 2004 December~4 and 8.  Each of the two spectrographs was used with a 1200B grating, giving a resolution of 2~\AA\ (dispersion 1.1~\AA~pixel$^{-1}$) over a wavelength range of 3800--4900~\AA.  The same system was used for the 2dFS survey in the SMC (Evans et al.\ 2004). 

AAOmega (AA$\Omega$) was an upgrade of the 2dF spectrographs to dual-beam systems with new fiber feeds, using the same front-end optics.  A pilot study in the LMC, led by J.Th.v.L.\ (with C.J.E.\ observing, hence ``Evans prev.\ unpubl.'' in the table), was undertaken during 2006 February 
22--24; only data from the blue arm are discussed here.  On the first night 
fields were observed in 30~Dor and N11 with a 1700B grating and two central wavelengths (4100 and 4700~\AA), yielding complete blue coverage of 3765--4985~\AA, at a resolution of 1~\AA\ (dispersion 0.33~\AA~pixel$^{-1}$). On the second night the 30~Dor field was also observed with a 1500V grating at a central wavelength of 4375~\AA, delivering coverage of 3975--4755~\AA\ at a resolution of 1.25~\AA\ (dispersion $\sim$0.4~\AA~pixel$^{-1}$).

Unpublished observations by Moffat are from the Argus multifiber
spectrograph at the Cerro Tololo Inter-American Observatory (CTIO) 4~m
telescope, as further specified by Walborn et al.\ (2002b); the spectral
resolution is 2~\AA.

The {\it Hubble Space Telescope\/}/Faint Object Spectrograph data of Walborn et al.\ (1999) have a dispersion of 3~\AA\ per diode, with a factor of 4 oversampling.  They were smoothed by 3~pixels to a formal resolution of
2.25~\AA\ in that study; the representation of Brey~73--2A shown there has better fixed-pattern noise suppression than the one available here.  The Anglo Australian Observatory/Royal Greenwich Observatory spectrograph 
(with fiber feed) data of Walborn \& Blades (1997) have a resolution of 
2.2~\AA\, with relatively low S/N and heavy nebular-line contamination.

The observations by Fari\~na et al.\ (2009) are from the Las Campanas
Observatory 2.5~m telescope with the Wide Field Reimaging CCD Camera in a
multislit/grism configuration; the spectral resolution is 3.1~\AA.

\section{Results}

\subsection{Spectral Morphology}

Spectrograms of all these stars are displayed in Figures~1--3, in order of 
advancing spectral type.  Note that all have been uniformly smoothed and 
rebinned as specified in the Figure~1 caption.  All of the classifications 
have been either newly derived or reexamined for consistency in this work.  
Two salient characteristics of the Onfp class are immediately apparent.  
First, as in the smaller Galactic sample, there is a substantial range in the 
appearance of the peculiar He~II $\lambda$4686 profiles, from strong emission with a small absorption reversal, to a dominant absorption feature with weak emission wings; and intermediate cases occur as well.  Second, there is also a wide range of line broadening, with a predominance of large values, which affects both the absorption and emission lines equally again as in the Galactic counterparts.  These properties suggest rapid rotation, in which the region producing the emission lines participates, with the minority smaller values plausibly arising from high inclinations.

The absorption to emission progression in the He~II $\lambda$4686 line of
normal O-type spectra has been identified as a luminosity criterion (Walborn 1971, 1973, 2009; Walborn \& Fitzpatrick 1990).  Clearly it would be hazardous to apply that inference to the peculiar Onfp profiles. Conservatively, only spectra with a strong, dominant emission feature
have been given luminosity class I here, while no luminosity class
has been assigned otherwise; cf. $\lambda$~Cep vs. BD~$+60^{\circ}$~2522 in Walborn (1973).  The physical luminosities of these stars are
investigated below by means of the MC distance moduli.

The comparable broadening of the absorption and emission lines, including
N~III $\lambda\lambda$4634-4640-4642, is in keeping with the selective
nature of the latter (Walborn 2001) and the evidence that they are
predominantly of photospheric origin. 

Comments on several of the individual spectra and related information follow.  In several cases, they are relevant to subsequent discussion of possible origins of the peculiar profiles and their temporal variations.

\noindent AA$\Omega$~30Dor~333:  It is possible that this very early spectrum does not belong in the Onfp category, but rather $\lambda$4686 has a ``composite'' P~Cygni profile (with weaker emission blueward of the
absorption component) that is characteristic of these extreme spectral types 
(Walborn et al.\ 2002b).  However, it is included here for completeness, 
since it satisfies the formal Onfp definition, and there could be a physical 
relationship.  This star is 5--31 of Testor \& Niemela (1998) in the H~II
region N158 south of 30~Doradus, also discussed by Massey et al.\ (2000)
and Walborn et al.\ (2002b).  Its image is clearly extended, indicating a
multiple system to which the extremely bright absolute magnitude in
Table~2 refers.

\noindent Brey~73--2A: This star is a possible blue straggler in a compact 
cluster (Walborn et al.\ 1999) within the 30~Doradus~B association (Schild
\& Testor 1992), which might represent a clue to the origin of the Onfp 
class, as further discussed later.  (ST~1--28 and ST~1--93 are in the same
association, the former also in a subclustering.)

\noindent AV~80: The apparent broad N~IV $\lambda$4058 line in the optical
observation shown by Walborn et al.\ (2000), which led to the spectral-type 
range given there, is much weaker if present at all in the UVES data shown 
here.  Further observations are required to determine whether a real spectral 
variation is involved.  Walborn et al.\ (2000) also showed the ultraviolet 
spectrum of this star; it has strong wind features as expected for its spectral 
type.  Rapid rotation is not expected in evolved, massive stars with strong 
winds, which should be braked, as further discussed below.  A quantitative 
analysis of this spectrum was performed by Heap, Lanz, \& Hubeny (2006).

\noindent Parker~841: This star is in the 30~Doradus core cluster (Walborn \& Blades 1997), which may indicate a very young object, unlike the majority of the Onfp class.

\noindent ST~2--32 and ST~2--53: These two very similar spectra correspond
to objects within 30~Doradus~C (Schild \& Testor 1992); the latter is one
of the brightest stars in the region and is in a subclustering. 

\noindent 2dFL~52--50: The $\lambda$4686 emission is single peaked, so this 
is not an Onfp spectrum.  However, both absorption and emission lines of this
apparent supergiant are unusually broadened, and as will be seen below, 
some Onfp objects with multiple observations intermittently remove
themselves from the class, so this may well be a related object.

\noindent AA$\Omega$~30Dor~320: This star is 5--82 of Testor \& Niemela (1998).

\noindent HDE~269702: This highly unusual supergiant spectrum has narrow lines and is one of only two in the sample without the ``n'' qualifier in its type; it also has the lowest value of $v\sin{i}$ measured here (as described below). The very narrow, double $\lambda$4686 emission is unique to date, but again, the correlation between absorption- and emission-line widths is typical of the Onfp class.  Note that this observation is from UVES, and that this profile would likely not be detected in low-resolution data.

\noindent Fari\~na~35: This is the X-ray binary LMC~X-1 (references in 
Fari\~na et al.\ 2009), so it is likely that the composite $\lambda$4686 
profile arises at least partly from systemic phenomena and not solely from 
the O-star extended atmosphere.  Nevertheless, it is retained in the present 
sample because it was selected from the survey data without knowledge of its 
identity.  It is the other relatively narrow-lined spectrum in this sample.  
As shown by van der Meer et al.\ (2007--their online Figure~2), double-peaked $\lambda$4686 profiles are common at certain phases in X-ray binaries, including SMC~X-1 and LMC~X-4.  Evans et al.\ (2004) reported both double- and single-peaked $\lambda$4686 profiles in different observations of SMC~X-1; see also Lennon (1997).

\noindent AV~321: Another unusual high-luminosity, very broad-lined 
spectrum.  There is no ``f'' parameter in the spectral type because such is 
not used later than O8.5.  The weak emission wings at $\lambda$4686 may
appear difficult to discern at the scale displayed here, but they are
striking in the original high-resolution, high-S/N UVES data.  Again,
this profile would likely not be detected in low-resolution data.

\subsection{Absolute Magnitudes}

Absolute visual magnitudes of the sample stars have been calculated from 
the photometry, with the assumptions of distance moduli of 18.6 for the
LMC and 19.1 for the SMC, a value of $R=3.0$ for the ratio of total to 
selective extinction, and intrinsic colors of the spectral types as
listed by Walborn et al.\ (2002a).  The results are given in Table~2 and 
plotted in Figure~4 together with the luminosity-class calibration of Walborn
(1973), updated for the O2--O3 stars by Walborn et al.\ (2002b).  As shown 
by the latter and references cited therein, the O spectral types are
insensitive to metallicity since they depend upon the He ionization ratio, 
and there is no evidence for systematic differences among the visual 
absolute-magnitude calibrations of the Galactic and MC O spectral types, 
within the random uncertainties further discussed below.

It is immediately apparent that there is no absolute-magnitude
distinction between the objects assigned luminosity class I and the
others in this Onfp sample.  Indeed, the isolation of peculiar objects
is a prerequisite for the recognition of trends and correlations among
normal spectra.  

Of course, objects that appear ``too bright'' may be unresolved multiple
systems, particularly at the distance of the MCs (e.g., Walborn 
et al.\ 1999).  Inversely, for an assumed distance, the adoption of $R=3$ will 
yield an absolute magnitude that is too faint if the actual ratio is larger, 
which it may well be in and around H~II regions as is the case for the most 
discrepant class I objects N11-20, ST~1-28, and ST~1-93.  These issues can be investigated in the future with improved angular resolution or temporal 
monitoring, and more extensive photometric data and reddening analyses, 
respectively.  However, for the present the most straightforward interpretation of Figure~4 is that the Onfp phenomenon occurs at all luminosities.

\subsection{Rotational Velocities}

Although the resolution of many of the spectrograms is too low to attempt a 
secure measurement of the rotational broadening $v_{\rm (e)}\sin{i}$, we can make reasonably accurate determinations of the line-broadening parameter $v\sin{i}$, in all but six cases for which the resolution and/or S/N are inadequate for even the latter.  Although processes other than rotation may contribute to the line broadening (e.g., macroturbulence), we expect rotation to be the dominant contributor for the sample of mainly broad-lined stars considered here.

To estimate $v\sin{i}$ values, the OSTAR2002 model grid (Lanz \& Hubeny 2003) was numerically spun up at 10~km~sec$^{-1}$ steps, with a standard rotational convolution (e.g., Gray 1992).  A gray-atmosphere, linear limb-darkening coefficient of 0.6 was used; but comparison with full surface integrations, accounting for the detailed wavelength dependence of the non-linear limb darkening across the line profile (similarly to Howarth \& Smith 2001), shows no important differences from these results.  Models with LMC abundances were used (Galactic models gave insignificant differences, and SMC models none).  For each observed spectrum, the rotationally broadened grids were then further convolved with a Gaussian matching the instrumental resolution, and the best-fit model (in $T_{\rm eff}$, $\log{g}$, and $v_{\rm (e)}\sin{i}$) identified as that giving the minimum r.m.s. residual over a fixed wavelength range.

Final results for $v\sin{i}$, included in Table~2, were determined using the 4250--4600~\AA\ range, in order to give the temperature-sensitive He lines more leverage with respect to the Balmer lines.  In practice, the measurements are reasonably insensitive to $T_{\rm eff}$ and $\log{g}$, but there is a systematic tendency for the best fits to occur at the lowest model gravities for a given $T_{\rm eff}$. The generally large $v\sin{i}$ results are not an artifact of data quality, as some of the largest values are derived from the high-resolution UVES data, and lower values are found among the low-resolution spectra.

Comparison of our $v\sin{i}$ values with those determined by Penny \& Gies 
(2009) from {\it Far Ultraviolet Spectroscopic Explorer} data for six stars 
in common shows reasonable agreement, with a negligible systematic offset and a median absolute difference of 30~km~sec$^{-1}$; this number is likely a reasonable upper limit to the uncertainties in our measurements.

Figure~5 compares the cumulative distribution function of $v\sin{i}$ for the 
MC Onfp sample (plus the ``slow'' rotator HDE~269702) with those for Galactic dwarfs and supergiants of spectral types O8 and earlier, that have 
determinations by Howarth et al.\ (1997) from {\it International Ultraviolet 
Explorer\/} data and empirical templates.  The Onfp sample differs from both of the latter with $>$99.99\% confidence according to a Kolmogorov-Smirnov test, that is, it is comprised of broader-lined spectra.  Again, we note that the Onfp class is defined by the He~II $\lambda$4686 profiles, while the line broadening is an independent variable.  Penny \& Gies (2009) have found that rotational velocity distributions are very similar between unevolved and evolved stars in the Galaxy, the LMC, and the SMC, as well as among the unevolved stars of the three galaxies.  On the other hand, they find evidence that macroturbulent broadening in the evolved stars is {\it lower\/} in the LMC and SMC samples than in the Galaxy.  Thus, the large offset between the MC Onfp and the Galactic spectra in Figure~5 is highly significant. 

\subsection{Spectral Variations}

Two observations each of N11--20 and Sk~$-67^{\circ}$~111 are shown in
Figures~6 and 7, respectively.  The N11--20 ESO/VLT/FLAMES data are from
the work of Evans et al.\ (2006) and were obtained on two consecutive
nights.  They display a significant change in the profile of He~II
$\lambda$4686.  From further radial-velocity and profile variations in
these data, Evans et al.\ concluded that N11--20 is a binary.

The change in the $\lambda$4686 profile of Sk~$-67^{\circ}$~111 between
two observations nine months apart is even more dramatic.  While the
previously unpublished Australian National University (ANU) 2.3~m data 
have lower resolution and S/N in the continuum, there is no doubt that 
the strong $\lambda$4686 emission line is single peaked, i.e., this is 
not an Onfp spectrum!  Prior evidence of variability in this spectrum 
is discussed by Walborn et al.\ (2002a); the ANU 2.3~m data parameters 
are given by Crowther et al.\ (2002).

The CTIO/Argus observations of Moffat et al.\ (unpublished) provided extensive temporal coverage specifically to search for spectroscopic binaries.
Nightly means from five consecutive nights of ST~1--28 and ST~1--93 are
displayed in Figure~8.  Large night-to-night variations in the $\lambda$4686 profiles are seen, again with non-Onfp single emission peaks on some nights.  Analysis of these data by G.~Skalkowski (unpublished) has revealed that ST~1--28 is a spectroscopic binary with a period of 2.35~d and a velocity full amplitude of 200~km~sec$^{-1}$, while ST~1--93 also displays large radial-velocity variations of undetermined period.

These scattered results strongly suggest that systematic monitoring of Onfp spectra is required to understand their nature.  In every case with more than one observation, significant variations are found.  An initial program on several of these objects was conducted by N.R.W.\ and A.~Ahumada
during October 2008 and results will be presented in a subsequent publication.

\subsection{Spatial Distribution}

Of course, all of the Onfp spectra correspond to young fields, but most of 
them are found at the peripheries of clusters and associations, or are even 
not obviously associated at all.  As noted in Section 3.1, only three of 
them are in bright H~II regions or tight clusters (AA$\Omega$~30Dor~333, 
Brey~73--2A, Parker~841) and two more are in looser subclusterings (ST~1--28, ST~2--53).  Interestingly, all five of these are among the earliest spectral types in the sample (Figure~1).  That is, the locations of the Onfp class appear analogous to those of normal evolving stars of similar spectral types, and they are not extremely young objects.  These results have implications for their origins and nature.

Alternatively, it could be relevant in this context that both $\zeta$~Pup 
and $\lambda$~Cep have been determined to be runaways (Mason et al.\  1998; Hoogerwerf, de Bruijne, \& de Zeeuw 2001; and references therein).  If that were a common property among the Onfp class, it would be related to their observed spatial distribution in the MCs.  Binary mass transfer, spinup of the mass gainer, supernova kicks, and gamma-ray bursts have all been associated with runaways in some models; thus, all of these topics could 
be interrelated in the Onfp class as further discussed below.

\section{Discussion: Possible Origins and Destiny of the Onfp Class}

\subsection{Meaning of the Onfp Category}

As a prologue to the discussion, it may be useful to re-emphasize that ``Onfp'' is a description of a (peculiar) spectral morphology, namely reversed He~II $\lambda$4686 emission profiles.  In most but not all cases, the latter feature is correlated with line broading indicative of rapid rotation.  In view of this definition, the observed spectral variations that at least in some cases temporarily remove objects from the category, and the {\it a priori\/} possibility that different physical situations may give rise to this general type of $\lambda$4686 profile, the term ``Onfp stars'' should be avoided, to prevent any misconception that the spectroscopic category corresponds to a unique, identified physical phenomenon.  There may be a predominant mechanism giving rise to the majority of the class, but it remains to be determined whether or not that is true.   

\subsection{Binaries}

In a double-lined O-type spectroscopic binary, it is in principle possible 
for (variable) He~II $\lambda$4686 profiles such as those in at least some
Onfp spectra to result from two emission features, or an emission line from 
one star and an absorption line from the other, shifting relative to each
other throughout the orbit.  In addition, there could be variable emission 
features arising from mass transfer and/or colliding winds in the system.  
As reported here, ST~1--28 is a spectroscopic binary, and the Galactic Onfp 
star HD~152248 is a well-studied, double-lined spectroscopic binary (Sana,
Rauw, \& Gosset 2001; Mayer et al.\ 2008).  In particular, Sana et al.\ showed 
that in HD~152248, both stellar $\lambda$4686 features are absorption lines 
and the emission most likely arises from the colliding winds.  However, its Onfp classification has been retained here because it was derived from a 
single, low-resolution observation (Walborn 1972), as is the case for many 
of the present MC sample, so the latter may contain some analogous objects. 
As also mentioned above, Fari\~na~35 is an X-ray binary, while ST~1--93
and N11--20 are suspected spectroscopic binaries.

Nevertheless, the general appearance of most Onfp spectra, including several
observed at very high resolution, with their comparable broadenings of both
absorption and emission lines, is not that of double-lined spectroscopic
binaries.  Of course, only one observation is available for most of them,
and clearly this issue can be resolved only by spectroscopic (and
photometric) monitoring.  It is possible that multiple mechanisms will
apply to some different members of the Onfp class, but it appears likely
that a dominant one is relevant to the majority.

\subsection{Rotation}

Characteristically, most of the Onfp class have broadened absorption and
emission lines, as shown in the figures and investigated in some detail
in Section 3.3.  While detailed profile fitting at high resolution and
S/N will be required to establish it, nevertheless rotation is strongly
suggested as the line broadening mechanism.  The range of $v\sin{i}$ 
values found could be intrinsic or caused by inclination effects; the 
predominance of abnormally large values suggests the latter.

The defining Onfp He~II $\lambda$4686 profiles are reminiscent of the
Balmer profiles of Be stars, which naturally suggested the possibility of
(hotter) disks in the former.  Evidence for line polarization effects in 
three Galactic Onfp spectra has been found by Harries, Howarth, \& Evans 
(2002) and Vink et al.\ (2009), but the interpretation is complex and any
implications for the presence of disks are uncertain; in fact, one of these 
cases is the binary HD~152248.  Moreover, Bouret et al.\ (2008) have reproduced variable, double-peaked emission profiles in one of these stars, $\zeta$~Pup, with a rotating, clumped wind and no disk.

\subsection{Mergers}

The persistence of rapid rotation in an evolved O-type star with a strong
wind presents a problem, since the wind should brake the rotation, as 
discussed by, e.g., Ekstr\"om et al.\ (2008) in addressing the greater 
relative frequency of rapid rotators and the Be phenomenon among the B stars.  Rapid rotation at a late age may be induced by mass transfer (Langer et al.\ 2008) and/or a merger in a binary system, in which orbital angular 
momentum is converted to rotation (e.g., Dale \& Davies 2006); such an 
object may also be a candidate gamma-ray burst progenitor, as they discuss.  
In this regard, it is interesting that Brey~73--2A lies about 2 mag above 
the apparent main-sequence turnoff in its compact cluster (Walborn et al.\ 
1999); it could be a blue straggler resulting from a binary merger.  Of course, this is a single case, and it could also be simply a result of unresolved multiplicity, although ST~2--53 may be a similar case.  The possibility of binary mass transfer and stellar mergers warrants consideration in further analysis of the Onfp phenomenon.

\section{Summary}

We have investigated a sample of 28 Onfp (and related) spectra in the 
Magellanic Clouds, defined primarily by composite, non-P-Cygni emission 
plus absorption profiles in the key He~II $\lambda$4686 line.  Their diverse spectral morphology, absolute magnitudes, line broadening, spectral variations, and spatial distribution have been surveyed. The salient results in each area are as follows.  There is a range of $\lambda$4686 profile shapes, but the broadening of emission and absorption lines is uniformly correlated.  There is no relationship between the emission-line strengths and the absolute magnitudes in these peculiar objects, as is the case for normal stars.  Large, likely rotational line broadening predominates, although a few narrow-lined objects are found; cumulative line-broadening distributions show
definitively that the Onfp spectra are drawn from a different parent population than normal spectra.  All objects with multiple observations
display significant spectral variations, which can even remove some spectra from the Onfp class, i.e., the $\lambda$4686 emission becomes single peaked, temporarily.  Only a few of these objects are located in compact clusters; most lie in the peripheries of associations or even in the (young) field.

Some of the Onfp objects are spectroscopic binaries, including at least one
X-ray binary, which may be the cause of the peculiar and variable $\lambda$4686 profiles in those cases; the current data are inadequate to determine
binarity or otherwise for most of them.  The most likely hypothesis for
the line broadening is rotational, but there is currently no compelling 
evidence for disks (as in Be stars) as the predominant source of these 
$\lambda$4686 profiles; in one Galactic case it has been reproduced by a 
rotating, clumped wind model.  Mass-transfer binaries or stellar mergers are 
possible origins of these evolved rapid rotators with strong winds, and they 
merit consideration as gamma-ray burst progenitors.

These results suggest several observational and analytical developments that 
will contribute to understanding the nature of the Onfp class.  Clearly,
intensive spectral and photometric monitoring, including radial-velocity
measurements, are essential to determine the nature of the endemic
variability seen in the limited available coverage.  High spectral
resolution and S/N will be preferable.  High spatial-resolution images to
search for multiplicity, and more extensive photometric and reddening
analyses, will improve the accuracy of the absolute magnitudes (although
a gap between spatially and kinematically determined multiplicities in
the Magellanic Clouds will remain for the immediate future).
Quantitative spectral analyses of high-quality data will establish (or
otherwise) the rotational interpretation of the line broadening and
derive chemical abundances (especially of CNO) and other parameters 
essential for a definitive physical model (or models) of the Onfp
spectral category.

\acknowledgments
We thank the AAO, ESO, CTIO, and LCO staffs for their support in the
acquisition of the data reported here.  Publication support was provided 
by the STScI Director's Discretionary Research Fund.  A.F.J.M.\ and N.S.L.\  
thank the National Sciences and Engineering Research Council of Canada 
and the Fonds Quebecois de la Recherche sur la Nature et les Technologies 
for financial assistance.  R.H.B.\ acknowledges partial support from
Universidad de La Serena Project DIULS CD08102.  We thank the anonymous
referee for some useful comments.

\clearpage
\begin{deluxetable}{llcrcl}
\rotate
\tablenum{1}
\tablewidth{0pt}
\tablecolumns{6}
\tablecaption{Galactic Onfp Spectra}
\tablehead{
\colhead{Name} &\colhead{Sp Type} &\colhead{$V$} &\colhead{$B-V$}
&\colhead{$v\sin i$} &\colhead{References}\\
&&&&[km~sec$^{-1}$]
}
\startdata
$\zeta$ Pup               &O4 I(n)fp         &2.25 &$-$0.28  &219
&Conti \& Niemela 1976\\

$\lambda$ Cep           &O6 I(n)fp       &5.05 &0.24      &219
&Conti \& Frost 1974\\

HD 14434                   &O5.5 Vn((f))p &8.49 &0.17      &423\rlap{:}
&Kendall et al. 1995, 1996; De Becker \& Rauw 2004\\

HD 14442                   &O5n(f)p         &9.22 &0.41      &273
&Kendall et al. 1995, 1996; De Becker \& Rauw 2004\\

HD 152248                  &O7 Ib:(n)(f)p  &6.13 &0.14     &159+165
&Sana et al.\ 2001, Mayer et al.\ 2008\\

HD 172175                  &O6 I(n)fp       &9.44 &0.64     &\nodata
&Walborn 1982, Sota et al.\ 2010\\

HD 192281                  &O5 Vn((f))p    &7.55 &0.38     &270
&De Becker \& Rauw 2004\\

BD $+60^{\circ}$ 2522 &O6.5(n)(f)p    &8.66  &0.40    &178--240     &Rauw
et al.\ 2003
\enddata
\end{deluxetable}

\clearpage

\begin{deluxetable}{lllllrclcl}
\tabletypesize{\small}
\rotate
\tablenum{2}
\tablewidth{0pt}
\tablecolumns{10}
\tablecaption{Onfp and Related Spectra in the Magellanic Clouds}

\tablehead{
\colhead{Name\tablenotemark{a}} &\colhead{RA (2000)} &\colhead{Dec (2000)}  
&\colhead{Sp Type} &\colhead{$V$} &\colhead{$B-V$} 
&\colhead{Ref\tablenotemark{b}} &\colhead{$M_V$} &\colhead{$v\sin i$}
&\colhead{Spectroscopic Reference}\\
&&&&&&&&[km~sec$^{-1}$]
}
\startdata
AV 80       &0 50 43.8 &$-$72 47 41  &O4-6n(f)p     &13.33 &$-$0.14
&1 &$-$6.3  &370    &Walborn et al.\ 2000\\
AV 321     &1 02 57.1 &$-$72 08 09  &O9 IInp        &13.88 &$-$0.21
&1 &$-$5.5  &280    &Crowther prev.\ unpubl.\\
Sk 190      &1 31 28.0 &$-$73 22 14  &O7.5n(f)p     &13.59 &$-$0.22
&2 &$-$5.8  &320    &Crowther prev.\ unpubl.\\
2dFS 3975 &1 31 28.0 &$-$73 22 14  &O8n(f)p       &13.54 &$-$0.18 
&3 &$-$6.0 &340        &Evans et al.\ 2004\\
N11$-$20      &4 56 50.3 &$-$66 31 04  &O5 Inf+p     &13.18 &$-$0.22
&4 &$-$5.7  &260    &Evans et al.\ 2006\\
Sk $-69^{\circ}$ 50  &4 57 15.1 &$-$69 20 20  &O7(n)(f)p  &13.26:&$-$0.13
&5 &$-$5.9: &210  &Crowther prev.\ unpubl.\\
AA$\Omega$ N11$-$87    &4 58 44.9 &$-$66 12 15  &O6.5(n)fp  &13.12 &$-$0.30 
&6 &$-$5.5  &180  &Evans prev.\ unpubl.\\
2dFL 51$-$106  &5 18 49.5 &$-$69 14 06  &O7(n)fp       &12.84 &$-$0.14
&7 &$-$6.3  &140    &Howarth prev.\ unpubl.\\
Sk $-65^{\circ}$ 47  &5 20 54.7 &$-$65 27 18  &O4 I(n)f+p    &12.51 &$-$0.18 
&8 &$-$6.5  &200  &Crowther prev.\ unpubl.\\
Sk $-67^{\circ}$ 111 &5 26 48.1 &$-$67 29 30  &O6 Ia(n)fpv   &12.57 &$-$0.20 
&8 &$-$6.4  &230  &Walborn et al.\ 2002a\\
2dFL 52$-$50   &5 29 01.7 &$-$68 32 04  &O7 Ianf       &13.26 &$-$0.15
&6 &$-$5.9  &230    &Howarth prev.\ unpubl.\\
2dFL 51$-$50   &5 31 08.1 &$-$68 36 54  &O7.5(n)(f)p   &12.82 &$-$0.18
&6 &$-$6.2  &270    &Howarth prev.\ unpubl.\\
HDE 269702  &5 31 52.1 &$-$67 34 21  &O8 I(f)p      &12.08 &$-$0.17
&8 &$-$6.9  &110    &Crowther prev.\ unpubl.\\
2dFL 52$-$171  &5 34 06.3 &$-$69 25 09  &O7(n)(f)p     &13.24 &$-$0.15
&6 &$-$5.9  &220    &Howarth prev.\ unpubl.\\
AA$\Omega$ 30Dor 142    &5 34 06.3 &$-$69 25 09  &O6.5(n)(f)p   &13.24 &$-$0.15
&6 &$-$5.9 &210  &Evans prev.\ unpubl.\\
AA$\Omega$ 30Dor 187    &5 35 51.9 &$-$69 23 19  &O6n(f)p       &13.40 &$-$0.18
&6 &$-$5.6  &350  &Evans prev.\ unpubl.\\
ST 2$-$32     &5 35 55.4 &$-$69 12 00  &O5n(f)p       &13.95  &0.09
&9 &$-$5.9  &\nodata    &Moffat prev.\ unpubl.\\
ST 2$-$53     &5 36 06.4 &$-$69 11 48  &O5n(f)p       &12.29 &$-$0.15
&9 &$-$6.8  &\nodata    &Moffat prev.\ unpubl.\\
ST 1$-$28     &5 37 38.0 &$-$69 10 15  &O3.5 Infp     &14.26  &0.18
&9 &$-$5.8  &\nodata    &Moffat prev.\ unpubl.\\
Brey 73$-$2A  &5 37 46.6 &$-$69 09 08  &O4(n)(f)p     &14.13  &0.13
&10 &$-$5.8  &\nodata    &Walborn et al.\ 1999\\
ST 1$-$93     &5 37 56.2 &$-$69 11 51  &O6 Ianfp      &14.88  &0.27
&9 &$-$5.5  &\nodata    &Moffat prev.\ unpubl.\\
Parker 841  &5 38 41.2 &$-$69 05 52  &O4$-$6(n)(f)p   &15.18  &0.17
&11 &$-$4.9  &\nodata    &Walborn \& Blades 1997\\
AA$\Omega$ 30Dor 320    &5 38 46.0 &$-$69 28 37  &O7n(f)p       &13.75  &0.05 &6 &$-$6.0  &300  &Evans prev.\ unpubl.\\
AA$\Omega$ 30Dor 333    &5 39 11.6 &$-$69 30 37  &O2$-$3(n)f*p    &12.27 &0.10 
&6 &$-$7.6  &170  &Evans prev.\ unpubl.\\
Fari\~na 35   &5 39 38.9 &$-$69 44 36  &O8(f)p        &14.61 &$-$0.08
&6 &$-$4.7 &\llap{$<$}100    &Fari\~na et al.\ 2009\\
Fari\~na 72   &5 39 59.8 &$-$69 36 11  &O6n(f+)p      &13.75 &$-$0.04
&6 &$-$5.7  &150    &Fari\~na et al.\ 2009\\
Fari\~na 82   &5 40 04.6 &$-$69 39 51  &O5n(f+)p      &12.38 &$-$0.03
&6 &$-$7.1  &280    &Fari\~na et al.\ 2009\\
AA$\Omega$ 30Dor 368    &5 40 13.8 &$-$69 25 35  &O7.5n(f)p     &13.30 &$-$0.12 &6 &$-$5.9  &280  &Evans prev.\ unpubl.\\
Fari\~na 151  &5 41 09.8 &$-$69 39 16  &O7n(f)p       &13.21  &0.19
&6 &$-$6.9  &270    &Fari\~na et al.\ 2009\\
AA$\Omega$ 30Dor 380    &5 41 11.2 &$-$69 55 44  &O7(n)(f)p     &13.44 &$-$0.03 
&6 &$-$6.0  &200  &Evans prev.\ unpubl.\\
\enddata
\tablenotetext{a}{AV: Azzopardi \& Vigneau 1982; Sk: Sanduleak 1969, 1970; 
N: Henize 1956;  AA$\Omega$: J.~van~Loon et al.\ unpublished survey; 
2dFL: I.~Howarth unpublished survey; ST: Schild \& Testor 1992; 
Brey: Breysacher et~al.\ 1999; Parker: Parker 1993.}
\tablenotetext{b}{Photometric References--1: Azzopardi \& Vigneau 1982;
2: Ardeberg 1980; 3: Massey 2002; 4: Evans et al.\ 2006; 5: Isserstedt 1979; 
6: Zaritsky et al.\ 2004; 7: Udalski et al.\ 2000; 8: Isserstedt 1975; 
9: Schild \& Testor 1992; 10: Walborn et al.\ 1999; 11: Parker 1993.}
\end{deluxetable}

\newpage

\begin{figure}
\epsscale{0.77}
\plotone{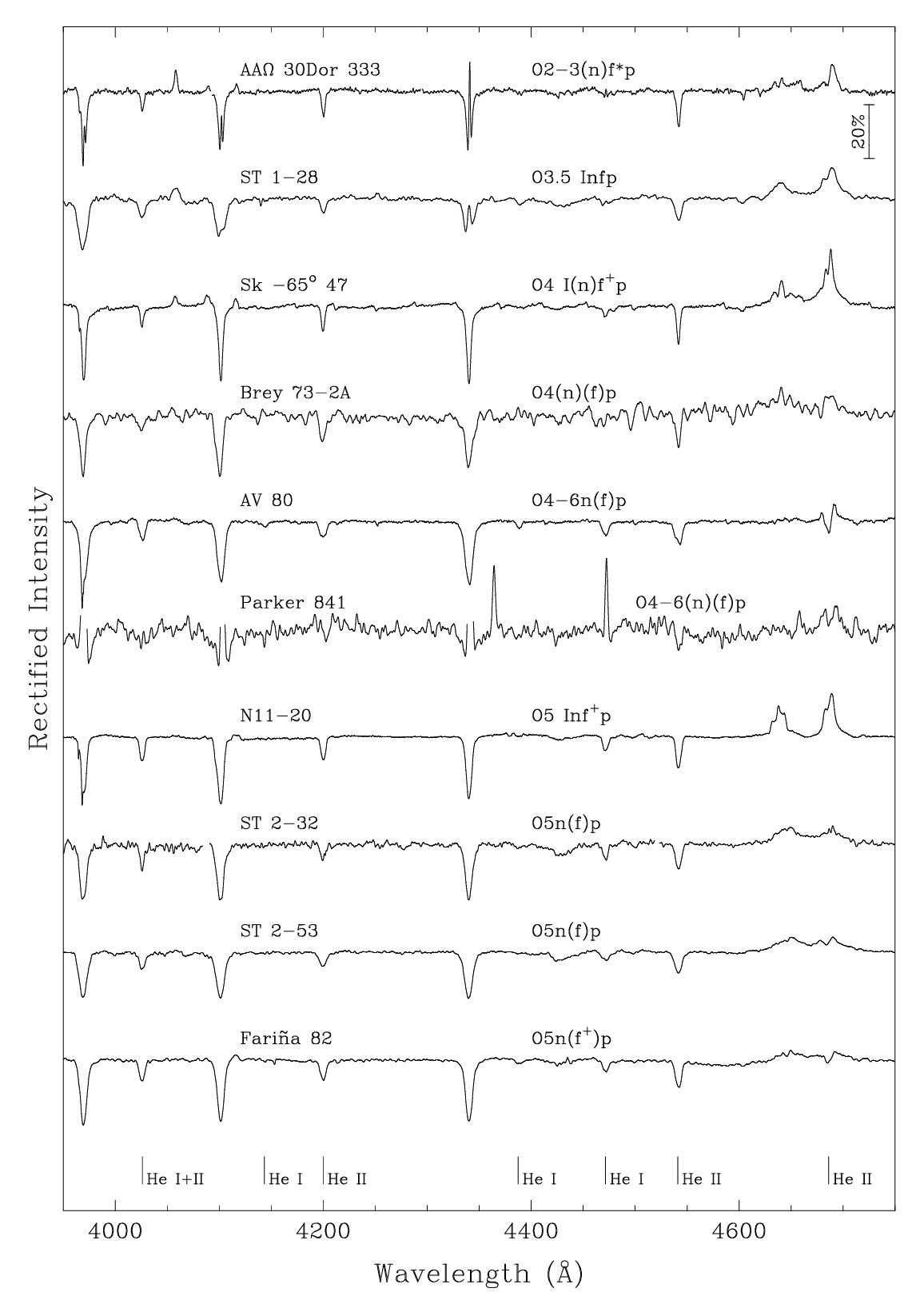}
\caption{\label{fig:fig1}
Rectified blue-green digital observations of MC Onfp spectra, in order of 
advancing spectral type.  The original data have a variety of resolutions; 
all have been smoothed with an 0.3~\AA\ Gaussian and rebinned here. 
Adjacent spectrograms are offset by 0.4 continuum unit; the vertical 
bar at upper right denotes 0.2 continuum unit.  Rest wavelengths are 
plotted.  The identified lines are He~I+II $\lambda$4026; He~I 
$\lambda\lambda$4144, 4387, 4471; He~II $\lambda\lambda$4200, 4541, 4686.}
\end{figure}

\begin{figure}
\epsscale{0.77}
\plotone{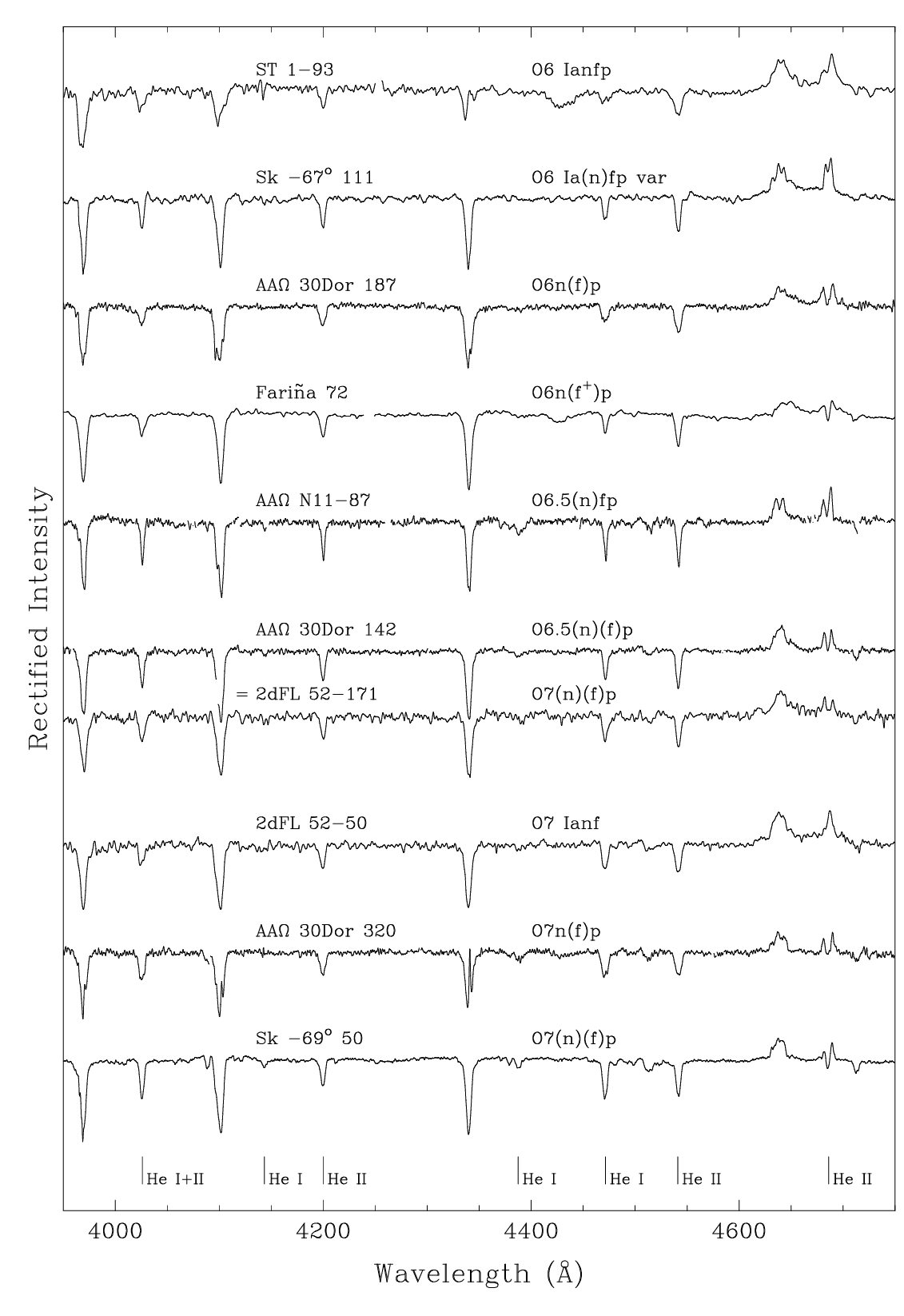}
\caption{\label{fig:fig2}
MC Onfp spectral-type sequence continued.  See the Fig.~1 caption.}
\end{figure}

\begin{figure}
\epsscale{0.77}
\plotone{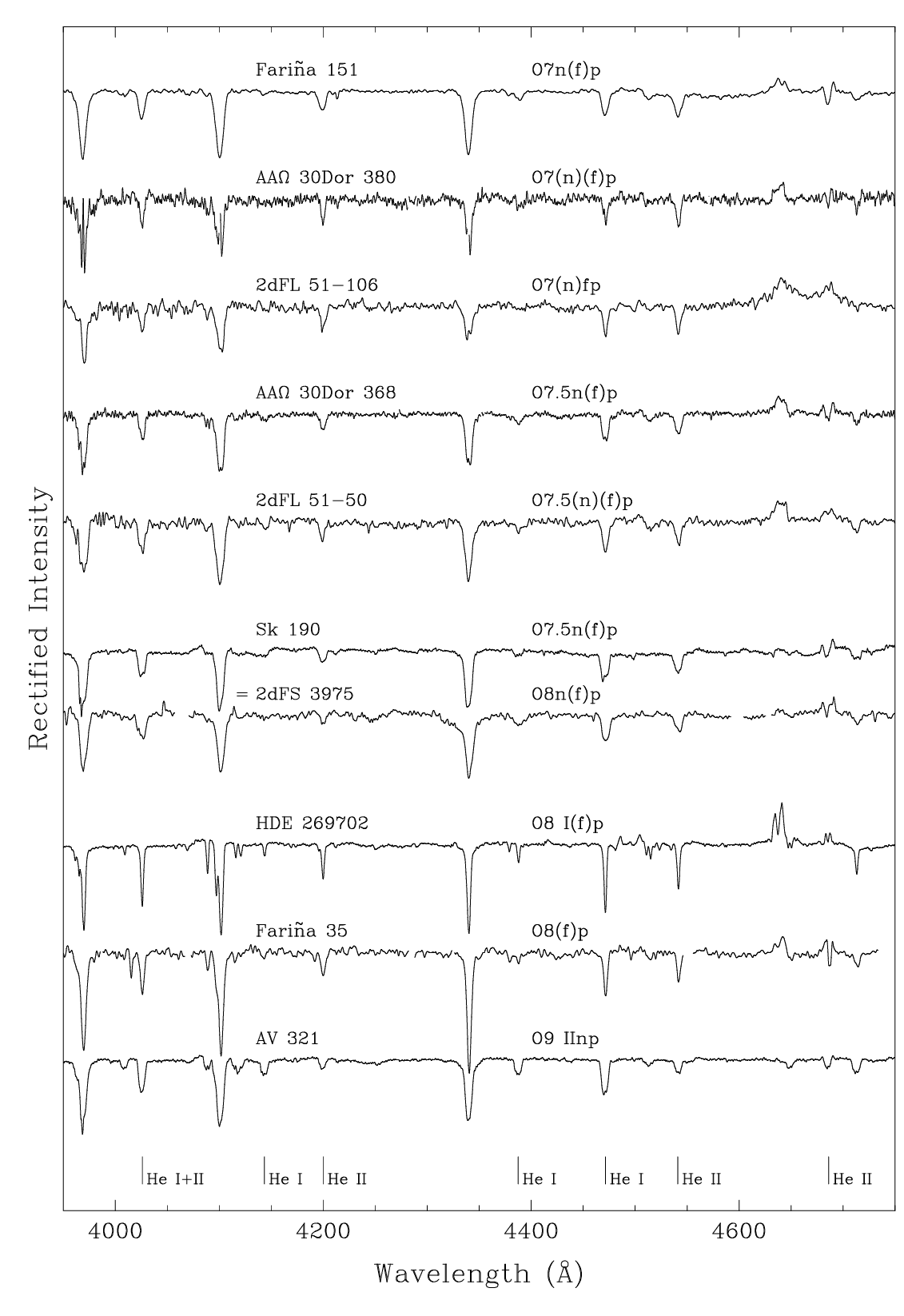}
\caption{\label{fig:fig3}
MC Onfp spectral-type sequence concluded.  See the Fig.~1 caption.}
\end{figure}

\begin{figure}
\includegraphics[height=.75\textheight,angle=-90]{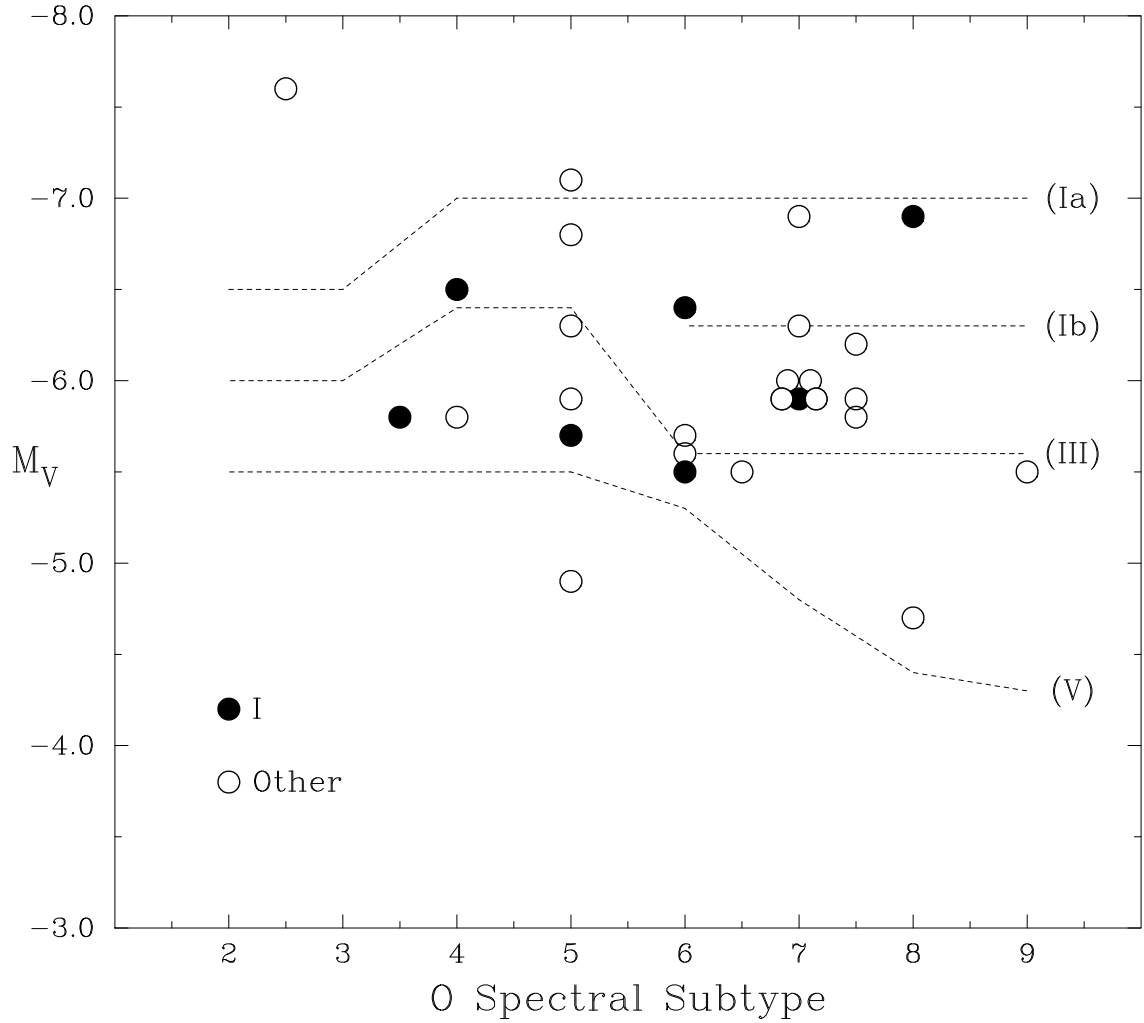}
\caption{
Absolute visual magnitudes of the Onfp sample plotted against spectral types.  
{\it Filled symbols\/}: luminosity class I; {\it open symbols\/}: no luminosity 
class assigned.  The dashed lines show the calibration for normal spectra by 
Walborn (1973), but Walborn et al.\ (2002b) for types O2--O3.}
\end{figure}

\begin{figure}
\includegraphics[height=.75\textheight,angle=-90]{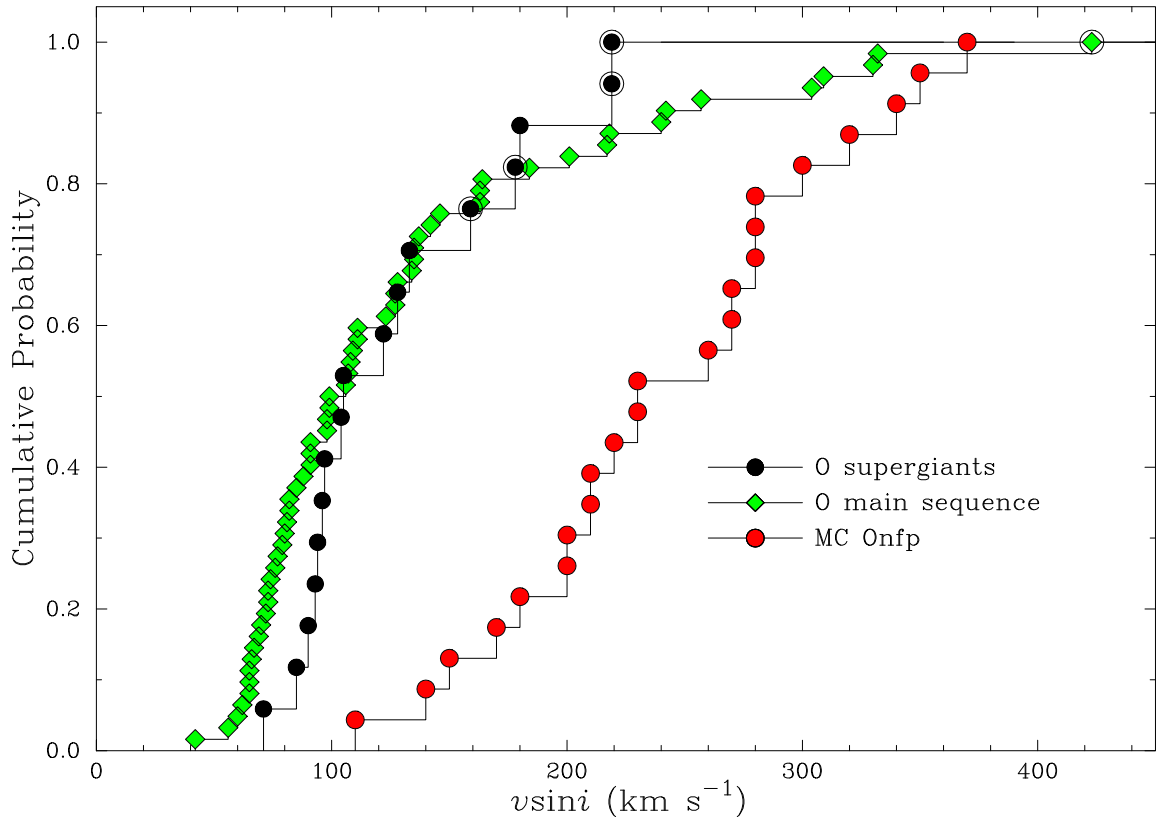}
\caption{
Cumulative probability functions of the line broadening $v\sin{i}$ for 
Galactic O supergiants and main sequence, compared to that of the MC Onfp 
sample (plus the slow rotator HDE~269702), with symbols as defined in the 
key; circled Galactic symbols are Onfp spectra.}
\end{figure}

\begin{figure}
\includegraphics[height=.75\textheight,angle=-90]{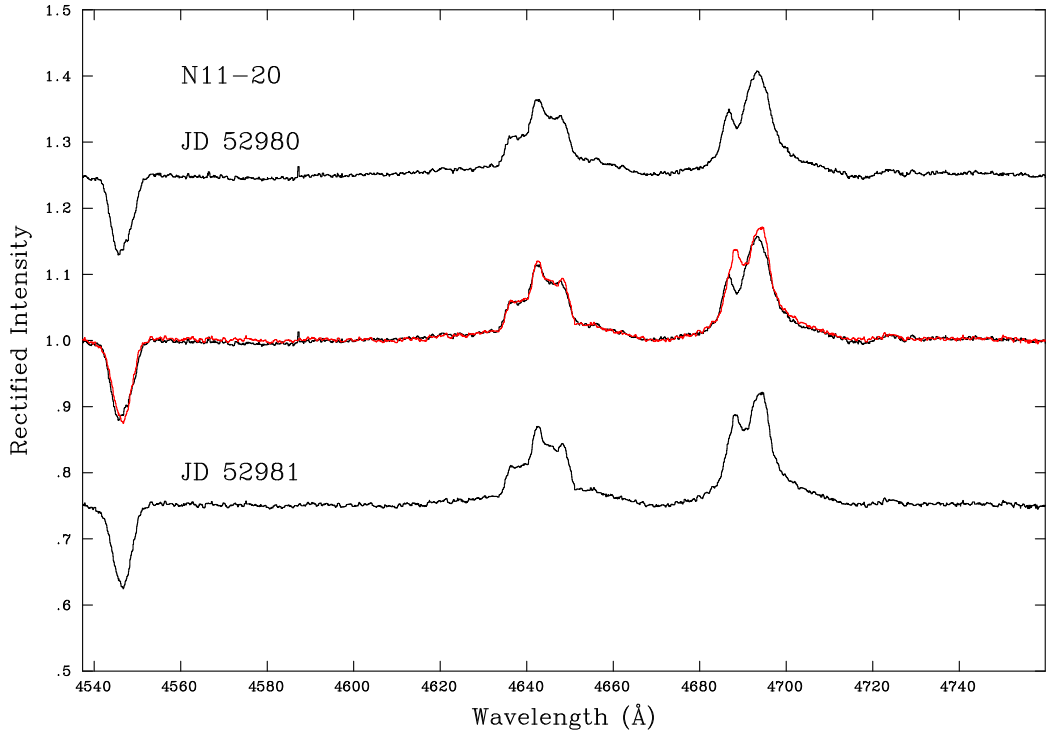}
\caption{
The green region in the spectrum of N11--20 on the two consecutive nights 
JD~24+ (top and bottom; superimposed in the middle plot).  In this and 
the subsequent figures, observed wavelengths are plotted.  The spectral 
features are He~II $\lambda$4541 absorption, a blend of the broadened 
N~III $\lambda\lambda$4634-4640-4642 emission lines, and the 
characteristic Onfp composite profile of He~II $\lambda$4686.  
The last of these varied significantly between the two nights.}  
\end{figure}

\begin{figure}
\includegraphics[height=.75\textheight,angle=-90]{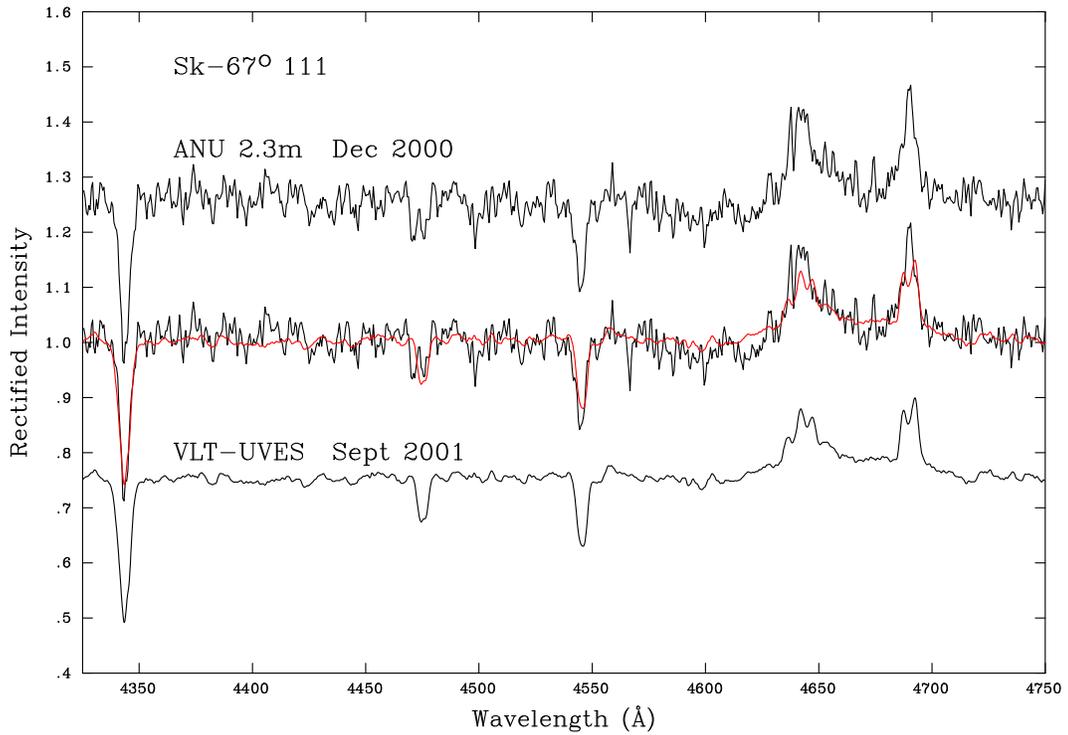}
\caption{
A similar plot to the preceding for Sk~$-67^{\circ}$~111, except that the
somewhat wider wavelength range also includes the H$\gamma$ $\lambda$4340
and He~I $\lambda$4471 absorption lines, and the two observations are
from different instruments used 9~months apart.  Despite the data
differences, the He~II $\lambda$4686 profile clearly changed in nature;
the single-peaked profile at the earlier epoch is not Onfp.}  
\end{figure}

\begin{figure}
\includegraphics[height=.8\textheight,angle=-90]{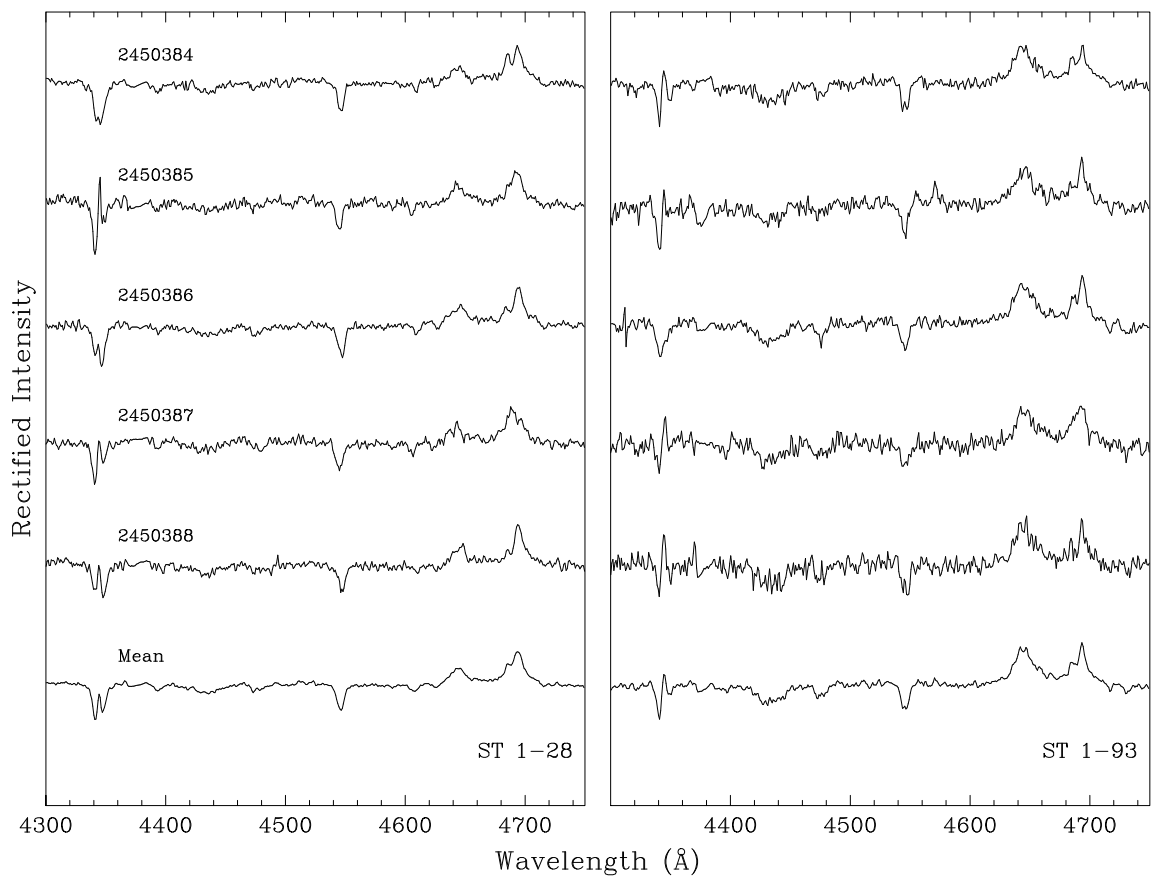}
\caption{
Observations in the same spectral range as the preceding figure of ST~1--28 
and ST~1--93 from 5 consecutive nights identified by JD; the offsets 
between adjacent spectrograms are 0.4 continuum unit.  Nebular H$\gamma$
emission and interstellar $\lambda$4430 are present.  There are significant 
night-to-night changes in the He~II $\lambda$4686 profiles.  As discussed 
in the text, both stars have variable radial velocities, which can be seen 
here in the relative shifts of the stellar and nebular H$\gamma$ lines.  
ST~1--28 has been found to be a spectroscopic binary with a period of 2.35~d; 
the $\lambda$4686 profile variations may be correlated.}
\end{figure}

\end{document}